\documentclass[epj-spec]{svjour}
\usepackage{graphics,graphicx}
\usepackage{subfigure}
\usepackage{amsfonts}
\usepackage{amsmath}
\usepackage{amssymb}
\usepackage{ulem}

\newcommand {\be} {\begin{equation}} 
\newcommand {\ee} {\end{equation}} 
\newcommand {\Be}{\begin{eqnarray*}}
\newcommand {\Ee} {\end{eqnarray*}}
\newcommand {\bey} {\begin{eqnarray}} 
\newcommand {\eey} {\end{eqnarray}} 

\begin{document}

\title{Anomalous transport and relaxation in 
classical one-dimensional models}

\author{G. Basile \inst{1} \and L. Delfini \inst{2} \and 
S. Lepri \inst{2}\fnmsep
\thanks{Corresponding Author, \email{stefano.lepri@isc.cnr.it}} 
\and R. Livi \inst{3} 
\thanks{Also at Istituto Nazionale di Fisica Nucleare 
and Istituto Nazionale di Fisica della Materia, Firenze.}
\and S. Olla \inst{4}
\and A. Politi \inst{2}
}                     
\institute{Dipartimento di Matematica, Universit\`a di Firenze, 
Viale Morgagni 67a, 50134 Firenze, Italy \and
Istituto dei Sistemi Complessi, Consiglio Nazionale
delle Ricerche, via Madonna del Piano 10, I-50019 Sesto Fiorentino, Italy.
\and  Dipartimento di Fisica, via G. Sansone 1 I-50019, Sesto
Fiorentino, Italy
\and Ceremade, UMR-CNRS 7534, Universit\'e de Paris Dauphine, 
75775 Paris Cedex 16, France
}

\date{Received: \today}

\abstract{After reviewing the main features of anomalous
energy transport in 1D systems, we report simulations performed with
chains of noisy anharmonic oscillators. The stochastic terms are added
in such a way to conserve total energy and momentum, thus keeping the 
basic hydrodynamic features of these models. The addition of this 
"conservative noise" allows to obtain a more efficient estimate of the 
power-law divergence of heat conductivity $\kappa(L) \sim L^{\alpha}$  
in the limit of small noise and large system size $L$.
By comparing the numerical results  
with rigorous predictions obtained for the harmonic chain, we show 
how finite--size and --time 
effects can be effectively controlled. For low noise amplitudes, the 
$\alpha$ values are close to 1/3 for asymmetric potentials and to 0.4 
for symmetric ones. These results support the previously conjectured 
two-universality-classes scenario.}

\maketitle  

\section{Introduction}

Many-body systems constrained in reduced spatial dimensions (1 and 2D) display
unusual relaxation and transport properties. Following familiar arguments of
equilibrium statistical mechanics, this should be traced back to the predominant
role of fluctuations that give rise to, e.g., long--range order etc. In 
the context of non--equilibrium processes, the presence of anomalies
is signalled by the appearance of long--time tails in the correlation
functions of the relevant currents \cite{PR75,kirk} leading to ill--defined 
transport coefficients i.e. to the breakdown of standard hydrodynamics. The
case of one--dimensional models is perhaps the most striking. For the sake of
an example, we mention the divergence of viscosity in cellular automata fluids
\cite{DLQ89}, anomalous diffusion in single-file systems ~\cite{singlef}, and
the enhancement of vibrational energy transmission in polymers~\cite{MHSU86} or
individual carbon nanotubes~\cite{maruyama}. 

Within this general context, one of the issues that attracted a renewed 
interest in the last decade is the problem of anomalous heat conduction in 
one-dimensional models (for a review see \cite{LLP03}). The interest in such
models originates from the need of constructing a minimal, nonperturbative 
theory 
of nonequilibrium stationary states and the quest for a rigorous microscopic
foundation of phenomenological relations (the Fourier's law in this case). This
motivation generated a vast literature, expecially in the mathematical physics
community \cite{BLR00}. Moreover, many of the peculiarities of 1D
models turned out to be of interest by themselves, as examples of highly
correlated, and thus complex, behaviour. Those latter features are shortly
reviewed in the first part of this paper. In the second part, the usefulness
of an additional stochastic noise \cite{olla} is discussed with reference to
some open questions.

The class of models we will consider consists
of a set of $N$ classical point--like particles with masses $m_n$ and 
positions $x_n$ ordered along a line. Interactions are restricted only to
nearest-neighbour pairs and the dynamics is ruled by 
\begin{equation} 
m_n{\ddot x}_n = - F_n + F_{n-1} \quad ; \qquad 
F_n=- V'(x_{n+1} - x_n)  \, ,
\label{eqmot} 
\end{equation}  
where $V'(y)$ is a shorthand notation for the first derivative of the
the interparticle potential $V$ with respect to the argument. Boundary
conditions for the first and last particle will be specified in the
various cases.
The two most intensively studied systems are the Fermi--Pasta--Ulam (FPU) model 
(with equal masses $m_n=m$) \cite{PRV67,JPW68,N70,LLP97}
\footnote{For the sake of nomenclature, it is useful to mention
two particular cases: the quadratic plus cubic ($g_4=0$) and quadratic plus 
quartic ($g_3=0$) potentials that, for historical reasons, are referred to as 
the ``$\alpha$-FPU'' and ``$\beta$-FPU'' models, respectively. In the former 
one, sufficiently small coupling constant $g_3$ and/or energies must be 
considered to avoid runaway instabilities.},
\begin{equation}
V(y)\;=\; \frac{g_2}{2}\,(y-a)^2+ \frac{g_3}{3}\,(y-a)^3 + 
\frac{g_4}{4}\, (y-a)^4 \, ,
\label{fpu}
\end{equation} 
and the diatomic Hard-Point Gas (HPG), where $m_{n}=m$ ($rm$) for even (odd)
$n$, with the interaction potential 
\cite{C86,H99,D01,GNY02}. 
\begin{displaymath}
 V(y)= \left \{ \begin{array}{ll}
     \infty & \quad \mbox{$y=0$}\\
     0 & \quad \mbox{otherwise}
     \end{array} \right.
\label{u}     
\end{displaymath}

We clarify from the beginning that we will always deal with genuine 
nonintegrable dynamics. For the FPU, this means working 
with high enough energies/temperatures to avoid all 
the difficulties induced by quasi-integrability and the 
associated slow relaxation to equilibrium \cite{chaos}. For the 
HPG this requires fixing a mass-ratio $r$ not too close to unity.

The results that emerged from a long series of works can be summarized
as follows. All studied models of the form (\ref{eqmot}) display  
\textit{anomalous transport and relaxation features}, meaning
by this that (at least) one of the following phenomena has 
been reported:
\begin{itemize}

\item The finite-size heat conductivity $\kappa(L)$  diverges as $L^\alpha$ in
the limit of a large system size $L\to \infty$ \cite{LLP97}. This means that
this transport coefficient is ill-defined in the thermodynamic limit, i.e. 
Fourier's law does not hold; 

\item The equilibrium correlator of the
energy current displays a nonintegrable power-law decay, 
$\langle J(t)J(0)\rangle \propto t^{-(1-\delta)}$ ($0\le\delta < 1$) 
for long times $t\to \infty$ \cite{LLP98}. Accordingly, 
the Green-Kubo formula yields an infinite value of the conductivity; 

\item Energy perturbations progate superdiffusively \cite{denis}: a local
perturbation of the energy broadens and its variance $\sigma^2(t)$ 
grows in time as $t^{\beta}$
with $\beta > 1$;

\item Relaxation of spontaneous fluctuations is fast 
(i.e. superexponential) \cite{Bishop2,sandri}: 
at variance with standard hydrodynamics, the typical
decay rate in time of fluctuations, $\tau(q)$, is found
to scale as $\tau(q) \sim q^{-z}$ (with $z<2$).
\end{itemize}

Altogether, these features indicate that the kinetics of energy carriers is so
correlated that they are able to propagate \textit{faster} with respect to the
the standard (diffusive) case.  In view of this common physical origin, it is
expected that the exponents describing such process will be related to each
other by some ``hyperscaling relations''. Their value should be
ultimately dictated by the dynamical scaling of the underlying dynamics.
Moreover, we may at least hope that they are largely independent of the
microscopic details, thus allowing for a classification  of anomalous behaviour
in terms of ``universality classes''.

Numerical studies ~\cite{LLP03} 
indicate that anomalies occur generically in 1 and 2D, whenever
\textit{momentum is conserved}. This is connected to the existence of
long-wavelength (Goldstone) modes  (an acoustic phonon branch in the linear
spectrum of (\ref{eqmot})) that are very weakly damped.
Indeed, it is sufficient to add external (e.g. substrate) forces to make
the anomaly disappear. This is precisely the case of the ding-a-ling
model~\cite{CFVV84} and of other models in the same class, like the
Frenkel--Kontorova \cite{GH85,HLZ98} or the nonlinear Klein-Gordon
chains \cite{AK00}. The only remarkable exception is the coupled
rotor chain ~\cite{GLPV00,GS00} where, however, different mechanisms are at
work.

Recently introduced stochastic models provide some mathematically 
rigorous results about the importance of momentum conservation. A
random exchange of momentum between neighbor particles is added to the
Hamitonian deterministic dynamics. These exchanges may conserve only the
energy \cite{BO05}, or also the total momentum \cite{olla}. While in
the first case conductivity remains finite and Fourier law is proved
for harmonic interaction \cite{BO05}, when total momentum is conserved
it diverges as $L^{1/2}$ \cite{olla}. In the anharmonic case,
conductivity is much harder to compute or estimate also with this
noise. So we performed some numerical simulations on the quartic and
cubic FPU models with momentum+energy conserving noise. 

In the first part of the paper we will review in more detail the anomalous
features mentioned above. The second part is devoted to some preliminary
numerical studies of the stochastic models.

\subsection{Diverging finite-size conductivity}

A natural way to simulate a heat conduction experiment consists in putting the
system in contact with two heat reservoirs operating at different temperatures
$T_+$ and $T_-$. Several models for the thermostats have been proposed
based on both deterministic 
and stochastic algorithms \cite{LLP03}. Regardless of the actual thermostatting
scheme, after a transient, an off-equilibrium stationary state sets in, with a
net heat current flowing through. The thermal conductivity $\kappa$ is then
estimated as the ratio between the time--averaged flux $\overline j$ and the
overall temperature gradient $(T_+-T_-)/L$, where  $L=aN$ is the chain length
($a$ denoting the lattice spacing). In this manner, $\kappa$
should be considered as an effective transport coefficient, accounting for 
both boundary and bulk scattering mechanisms. The average $\overline j$ can 
be estimated in several
equivalent ways, depending on the employed thermostatting scheme. One
possibility is to directly measure the energy exchanges with the two baths. A
more general definition (thermostat-independent) consists in  averaging 
\begin{equation}  
j_n = \frac{1}{2} ({\dot x}_{n+1} + {\dot
x}_n) \, F(x_{n+1} - x_n) \, , 
\label{hf2}
 \end{equation} which
is obtained from the continuity equation for the energy density 
\begin{equation}
  \begin{split}
    e_n &= \frac12 m_n{\dot x}_n^2 + \frac12 \bigg[ V(x_{n+1} - x_n) +
    V(x_n - x_{n-1}) \bigg]\\
    \dot e_n &= j_{n-1} - j_n
  \end{split}
\label{hamilton} 
\end{equation} 


As a result of many independent simulations performed with the above-described
methods, it is now established that $\kappa \propto N^\alpha$. It is also 
remarkable that the same type of behaviour has been observed for a realistic
model of a single--walled carbon nanotube \cite{maruyama}. This type of molecular
dynamics simulations, that involves complicated three-body interactions
for carbon atoms, confirm that toy models like ours can indeed capture 
some general features. 

\subsection{Long-time tails}

In the spirit of linear--response theory, transport coefficients can be
computed from equilibrium fluctuations of the associated currents. 
More precisely, by introducing the total heat flux
\begin{equation}
J = \sum_n j_n ,
\label{tot-heat}
\end{equation}
the Green-Kubo formalism tells us that heat conductivity is determined from 
the expression
\begin{equation}
\kappa \;=\; \frac{1}{k_BT^2}\lim_{t\to\infty} \lim_{N\to\infty} 
\frac{1}{N}\int_0^t 
 \, dt' \, \langle J(t')J(0) \rangle 
\label{GK}
\end{equation}
where the average is performed in a suitable equilibrium ensemble, e.g.
microcanonical with zero total momentum. 

A condition for the formula (\ref{GK}) to give a well--defined conductivity
is that the time integral is convergent. This is clearly not the case when the
current correlator vanishes as $\langle J(t)J(0)\rangle \propto
t^{-(1-\delta)}$ with $0\le \delta < 1$. Here, the integral diverges as
$t^\delta$ and we may thus define a finite--size conductivity $\kappa(L)$ by
truncating the time integral in the above equation to $t\approx L/c$, where $c$
is the sound velocity. Consistency with the above definition thus implies
$\alpha=\delta$. The available data are consistent with this expectation, thus
providing independent method for estimating the exponent $\alpha$.

For later purposes, we mention that, by means of the Wiener--Khintchine
theorem, one can equivalently extract $\delta$ from the low-frequency 
behaviour of the spectrum of current fluctuations 
\begin{equation}
S(\omega) \equiv \int \, d\omega \langle J(t)J(0)\rangle {\rm e}^{i\omega t}
\label{Fourier}
\end{equation}
that displays a low--frequency singularity of the form $S(\omega) \propto
\omega^{-\delta}$. From the practical point of view, this turns out to be the
most accurate numerical strategy as divergence rates are better estimated than 
vanishing ones.

\subsection{Diffusion of perturbations}

Consider the infinite system at equilibrium with
an energy $e_0$ per particle and average momentum 0,
 and perturb it by increasing 
the energy of a subset of adjacent particles
by some preassigned amount $\Delta e$~. Let us denote 
by $e(x,t)$ the energy profile evolving from such a perturbed 
initial condition (for simplicity we identify $x$ with the average particle 
location $n\ell$). We then ask how the perturbation 
\be
f(x,t) = \langle e(x,t) - e_0 \rangle
\label{fdef}
\ee
behaves in time and space \cite{helfand}, where the angular brackets 
denote an ensemble average 
over independent trajectories. Because of energy conservation,
$\sum_n f(n\ell,t) = \Delta e$  at all times, so that $f$ can be
interpreted as a probability density (provided that it is also
positive-defined and normalized). 

At sufficiently long times and for large $x$, 
one expects $f(x,t)$ to scale as
\begin{equation}
f(x,t)=t^{-\gamma}\mathcal G (x/t^{\gamma}) 
\label{eq:scail}
\end{equation}
for some probability distribution $\mathcal G$ and a parameter 
$0\le\gamma\le 1$.

The case $\gamma=1/2$ corresponds to normal
diffusion and to a normal conductivity. On the other
hand, $\gamma=1$ corresponds to a ballistic motion and to a linear
divergence of the conductivity.
 
Consequently an $\alpha$ strictly contained between 0 and 1 implies a
superdiffusive behaviour of the macroscopic evolution of an energy
perturbation. It is an open problem to determine the dynamical nature
of this superdiffusion. In  \cite{klaft} has been proposed that
\textit{L\`evy walks} \cite{blumen} may describe these dynamics.

\subsection{Relaxation of spontaneous fluctuations}

The evolution of a fluctuation of wavenumber $q$ excited at $t=0$ is 
described by its correlation functions $G(q,t)$ .
For 1D models like (\ref{eqmot}) they are defined by considering 
the relative displacements $u_n=x_n-n\ell$ and defining the
collective coordinates
\begin{equation}
Q(q) \;=\;\frac{1}{\sqrt{N}} \sum_{n=1}^N \, u_n \, \exp({-iqn})
\quad q=\frac{2\pi k}{Na} 
\label{Uk}
\end{equation}
for $k$ being an integer comprised between $-N/2+1$ and $N/2$.
The normalized correlator $G(q,t)= \langle Q^*(q,t)Q(q,0) \rangle/\langle
|Q(q)|^2 \rangle$, is thus proportional to the density--density 
correlator, which is routinely used in condensed--matter physics
\cite{forster}.  

For sufficiently small wavenumbers $q$, the Fourier transform of $G$, the
dynamical structure factor $S(q,\omega)$,  usually displays sharp peaks at
finite frequency. The associated linewidths are a measure of the fluctuation's
inverse lifetime. Simulations  \cite{chaos,sandri} indicate that these
lifetimes scale as $q^{-z}$ with $z=1.50 - 1.67$. As explained above, one may
think of this as a superdiffusive process, intermediate between standard
diffusive and  ballistic propagation. 

This property is supported by theoretical results obtained in the framework of
Mode-Coupling Theory (MCT). This approach has been traditionally invoked
to estimate long-time tails of fluids \cite{PR75} and to describe the glass
transition \cite{mct}.  Basically, it amounts to writing 
a set of approximate equations for $G(q,t)$ that must be solved 
self--consistently.
For the problem at hand, the simplest version of the theory amounts to
consider the equations ~\cite{SS97,L98}
\begin{equation}
\nonumber
{\ddot G} (q,t) + 
\varepsilon \int_0^t \Gamma (q,t-s) {\dot G}(q,s) \, ds 
+ {\omega}^2(q) G(q,t)  
= 0 \quad,
\label{mct}
\end{equation}
where the memory kernel $\Gamma(q,t)$ is proportional to $\langle
\mathcal{F}(q,t)\mathcal{F}(q,0) \rangle$, with $\mathcal{F}(q)$ being the
nonlinear part of the fluctuating force between particles. Equations (\ref{mct})
must be solved with the initial conditions $G(q,0)=1$  and $\dot G(q,0)=0$. 
Equations ~(\ref{mct}) are derived within the well--known
Mori--Zwanzig projection approach \cite{KT}. 

The mode--coupling approach basically amounts to replacing the exact memory
function $\Gamma$ with an approximate one, where higher--orders correlators are
written in terms of $G(q,t)$. 
In the generic case, in which $g_3$ is different from
zero, the lowest-order mode coupling approximation of the memory kernel turns
out to be \cite{SS97,L98} 
\begin{equation}
\Gamma(q,t)= \,\omega^{2}(q)
\,\frac{2 \pi}{N} \sum_{p+p'-q=0,\pm\pi}  \,G(p,t) G(p',t) \quad .
\label{mct2}
\end{equation}
Here $p$ and $p'$ range over the whole Brillouin zone (from $-\pi$
to $\pi$ in our units)~. This yields a closed system of nonlinear
integro--differential equations. Both the coupling constant
$\varepsilon$ and the frequency $\omega(q)$ are temperature-dependent input
parameters, which should be computed independently by numerical simulations or
approximate analytical estimates \cite{SS97,L98}. For the aims of the present
work, we may restrict ourselves to considering their bare values, obtained in
the harmonic approximation. In the adopted dimensionless units they read
$\varepsilon = {3g_3^2 k_BT / 2\pi}$ and  $\omega(q)=2 | \sin\frac{q}{2}|$. Of
course, the actual renormalized values are needed for a quantitative comparison
with specific models. 

The long-time behaviour of $G$ can be determined by looking for a solution of 
the form 
\begin{equation}
G(q,t) \;=\; C(q,t )e^{i \omega (q)t} + c.c. \quad .
\label{g}
\end{equation}
with $\dot G \ll \omega G$. It can thus be shown ~\cite{DLLP06,DLLP07} that, 
for small $q$-values and long times $C(q,t) = g(\sqrt{\varepsilon}t q^{3/2})$
i.e. $z=3/2$ in agreement with the above mentioned numerics. 
Furthermore, in the limit $\sqrt{\varepsilon}t q^{3/2} \to 0$ one can
explicitely evaluate the functional form of $g$, thus obtaining
\begin{equation}
C(q,t) \;=\; \frac12 \exp\left( -D q^2 |t|^{\frac43} \right) \quad,
\end{equation}
where $D$ is a suitable constant of order unity.
Correlation display a ``compressed exponential'' behaviour \cite{B07} 
in this time
range. This also means that the lineshapes of the structure factors
$S(q,\omega)$ are non-Lorenzian but rather display a faster power-law
decay $(\omega - \omega_{max})^{-7/3}$ around their maximum.  

Upon inserting this scaling result into the definition of the heat flux, one
eventually concludes that the conductivity $\kappa$ diverges with a rate
$\alpha=1/3$.

\section{Universality}

The crucial question at this point is: how universal are the above defined exponents?
The renormalization group argument by Narayan and Ramaswamy~\cite{NR02},
predicts $\alpha = (2-d)/(2+d)$. 
Following the arguments exposed above, this implies that
in 1D the values of the exponents are
\begin{equation}
\alpha \;=\; \delta \;=\; \frac13, 
\qquad \beta =\frac43,  \qquad z \;=\;\gamma^{-1} \;=\; \frac32 \quad.
\label{13}
\end{equation}
According to this approach, any possible additional term in the noisy
Navier-Stokes equation yields irrelevant corrections in the renormalization
procedure~\cite{NR02}. 

On the other hand, there now exists a rigorous result, proving that $\alpha=1/2$
\cite{olla} in a chain of harmonic oscillators subject to an energy and 
momentum--conserving noise. Moreover, the application of kinetic theories to the
$\beta$-FPU model \cite{P03,N07,S07} predict $\alpha=2/5$, while
a modified version of the MCT adapted to this specific case 
gives instead $\alpha=1/2$ \cite{canadesi,DLLP07}. 

The validation of these theoretical results by numerical simulations is, to some 
extent, challenging. Generally speaking, the
available numerical estimates of $\alpha$ and $\delta$ range between 0.25 and
0.44~\cite{LLP03,LLP03b}. The existence of crossovers among different scaling
regimes has been observed~\cite{WL04}. However, even in the most favorable cases of
computationally efficient models as the HPG, finite--size corrections to 
scaling are sizeable. As a matter of fact, $\alpha$--values as diverse as 0.33
~\cite{GNY02} and 0.25~\cite{CP03} for comparable parameter choices have been
reported. On the other hand, a numerically convincing confirmation of the 
$\alpha=1/3$ prediction comes from the diffusion of perturbations \cite{denis}.
The most compelling deviations from the values given in (\ref{13}) have been reported for
the $\beta$-FPU model \cite{LLP03b}, where a better agreement with the predictions
of the kinetic theory has been observed.
 
A reasonable argument that can be invoked to delimit the $\alpha=1/3$ universality class
appears to be the \textit{symmetry} of the interaction potential with respect to
the equilibrium position. In fact, systematically larger $\alpha$-values have been 
reported only for symmetric potentials such as the $\beta$-FPU model or,
more recently, for a modified version of the HPG model \cite{EPJST}, where the 
interaction can be tuned to yield symmetric fluctuations of the force. 
Moreover, the rigorous result $\alpha=1/2$ \cite{olla} indeed concerns a model
whose potential (harmonic) is symmetric and where the stochastic updating is symmetric 
as well.
In the framework of MCT, one can understand that the symmetry of the fluctuations implies
that the quadratic kernel (\ref{mct2}) should be replaced by a cubic one 
\cite{DLLP07}, thus yielding different values of the exponents (see \cite{canadesi} 
for a thermodynamic interpretation of this difference). 

Whether symmetry is the only necessary ingredient to identify the asymptotic
behaviour is however not fully proven.  As mentioned above, difficulties
manifestly arise e.g. for the $\beta$-FPU model where it is not yet completely
assessed which of the predicted exponents one should expect. In this respect
some further reconsideration of MCT and kinetic theories may be of help.
Actually, there are even some controversial results about the dynamical scaling
exponent $z$ in the supposed broad  $\alpha=1/3$ class \cite{klages}. 
It is thus clear that further rigorous results on simple models would be 
definitely welcome.

\section{Stochastic models with energy and momentum conservation}

Recently, some of the authors proposed a new class of models for analyzing
the anomalous properties of heat conduction for a system of oscillators
\cite{olla}. The Hamiltonian dynamics is perturbed by a stochastic noise,
which acts only on the momenta, while conserving total energy and 
momentum. In particular, such an approach allows to compute explicitly the
heat flux correlation function in the harmonic case, i.e.
model~(\ref{fpu})  with $g_3 = g_4 =0$. For the 1D model it is found that
the current correlator vanishes as $\langle J(t)J(0)\rangle \sim
t^{-1/2}$, thus implying that $\kappa \sim L^{1/2}$. 
The average of the current correlator is performed over the
equilibrium measure, which is the uniform measure over the hypersurface of
constant total energy and momentum. As usual in this microcanonical measure, the
total momentum is set to zero. There are various ways for translating such
stochastic dynamics into a suitable algorithm. The numerical simulations
hereafter reported have been performed by making the system evolve over a
finite number $n$ of steps (each of duration $h)$ and, then,
by updating the momenta of $n_c$ triplets of nearest-neighbor oscillators,  
whose position are randomly chosen with a uniform probability density over the
chain sites. Each ``collision'' in the momentum-subspace is performed by
extracting from the uniform distribution over $[0,2\pi]$ a random angle
which rotates the three momenta in such a way to maintain fixed their sum
and the sum of their squares. The resulting configuration 
is the initial condition for a new deterministic trajectory lasting
again over $n$ steps, and so forth. 
Because of the conservation laws, the microcanonical measure (uniform
on the energy surface) is still invariant and it can be proven to be
ergodic.

The algorithmic procedure introduces a new control parameter in the dynamics,
namely the ratio between the number of collisions per the unit time and the
number of particles $N$.

\begin{figure}[ht]
\begin{center}
\includegraphics[clip,width=6cm]{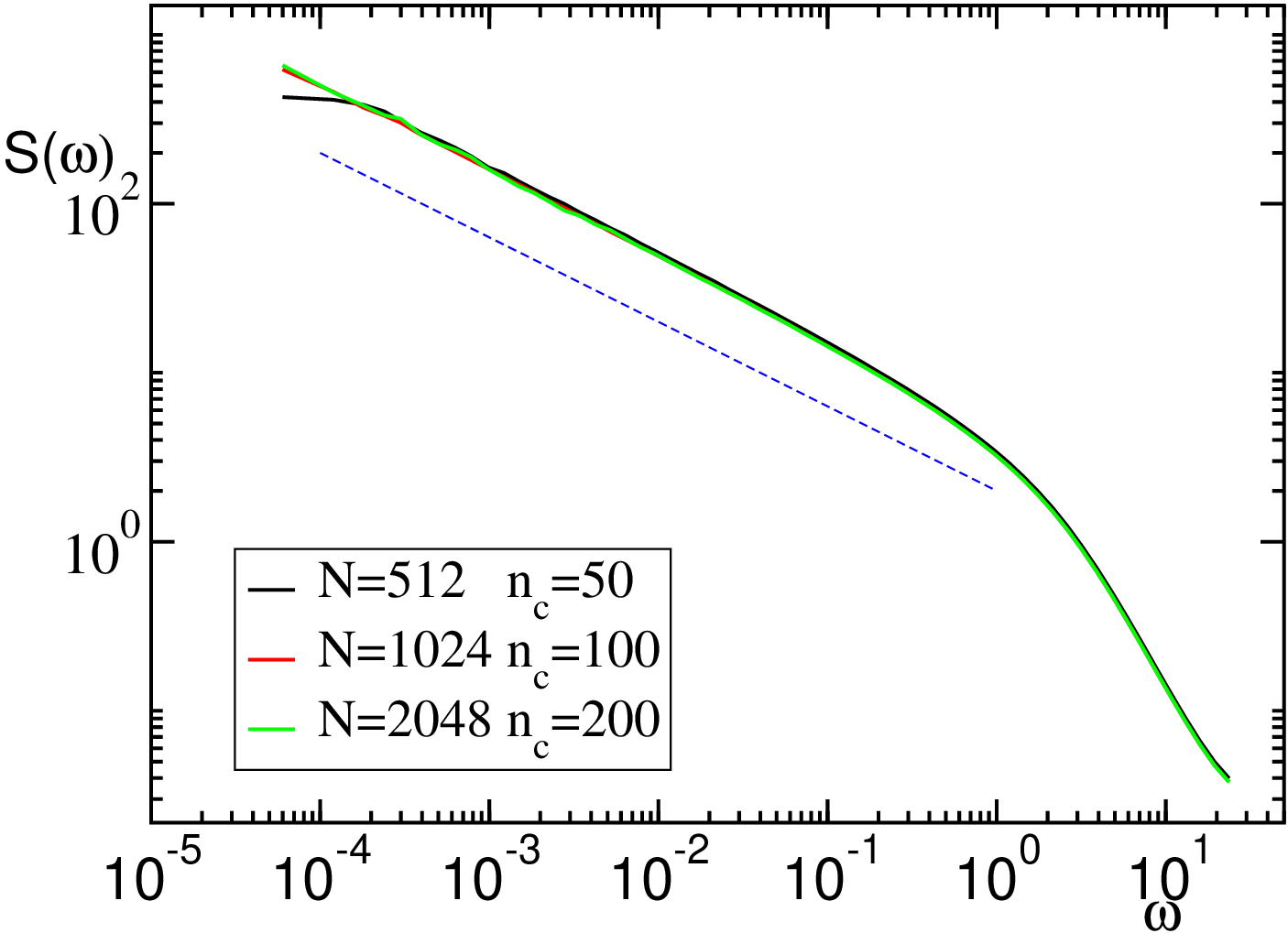}
\includegraphics[clip,width=6cm]{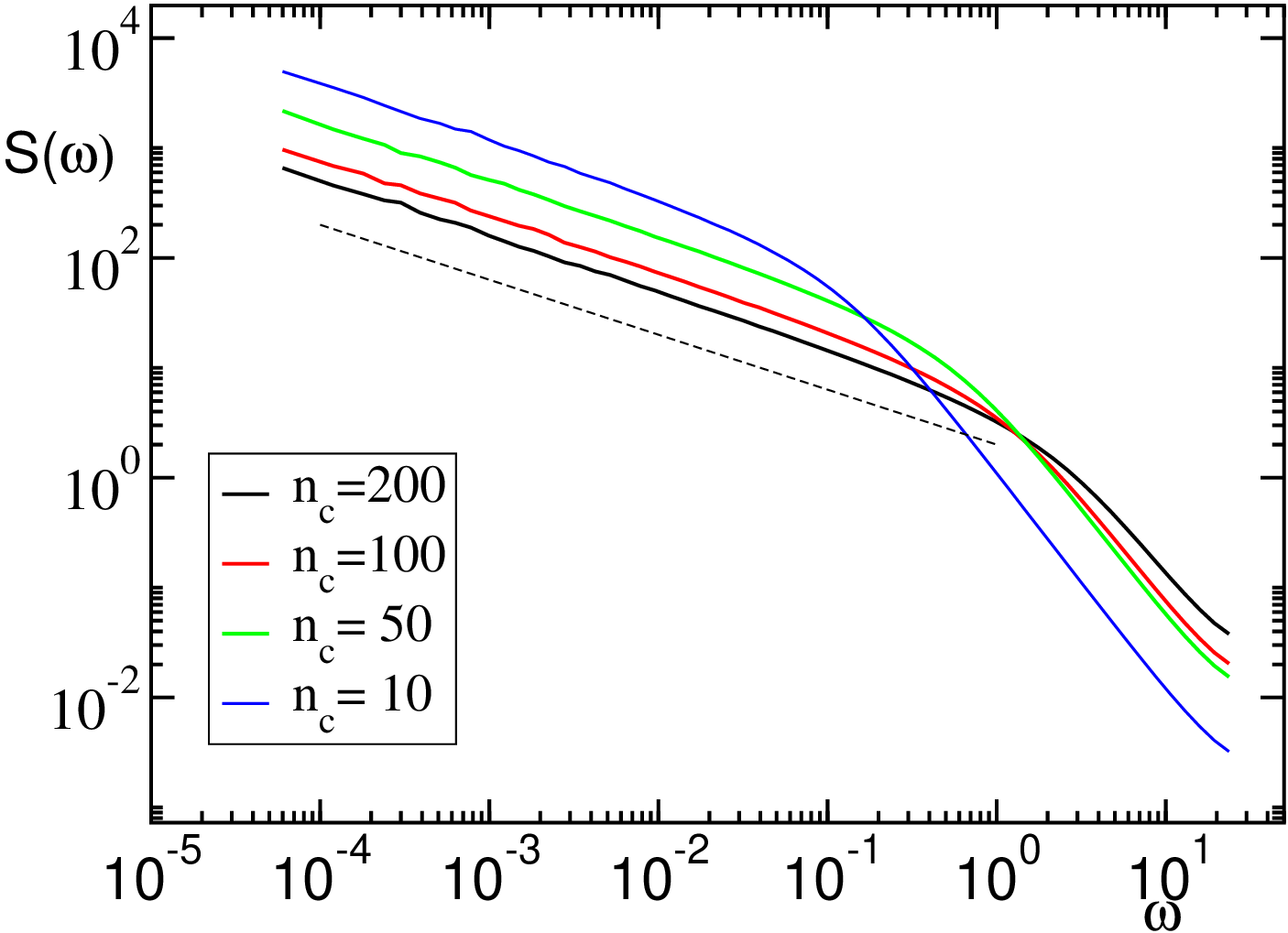}
\caption{Power spectra of the heat flux for the 
stochastic harmonic chain at energy density $e=10$. 
Units along the vertical axis are arbitrary.
Left panel: 
data for different values of $N$, while maintaining 
fixed the ratio $n_c/N$. 
Right panel: $N=2048$, and 
different values of $n_c$. Each curve is averaged over
about 500 independent trajectories. Here and in the following
figures,  
the time interval between collisions is $10\,h$ with integration
time step $h=0.01$. The thin dashed line corresponds to the law 
$\omega^{-1/2}$. } 
\label{fig1}
\end{center}
\end{figure}

In order to test the effectiveness of the algorithm, we have first
performed numerical simulations of the stochastic dynamics of the harmonic
chain, as we can make a comparison with the rigorous results \cite{olla}. 
In the left panel of figure \ref{fig1} we show the power spectrum
defined in (\ref{Fourier}) for different values of $N$, while maintaining
fixed the ratio $n_c/N \simeq 0.1 $. All curves align to the same
power--law behavior in a range of small values of $\omega$
of about 3 decades. In particular, the
spectrum is found to diverge as $S(\omega) \sim \omega^{-\delta}$ with
$\delta$ slightly larger than $1/2$. 
Moreover, finite size effects are clearly 
visible in the spectrum corresponding to the smallest chain  length
($N=512$). In the right panel of figure \ref{fig1}, we have kept $N=2048$,
while we have considered different values of $n_c$. For decreasing values
of $n_c$, we see that $\delta$ remains close to
$1/2$, in agreement with the rigorous prediction reported in \cite{olla}. 

\begin{figure}[ht]
\begin{center}
\includegraphics[clip,width=8cm]{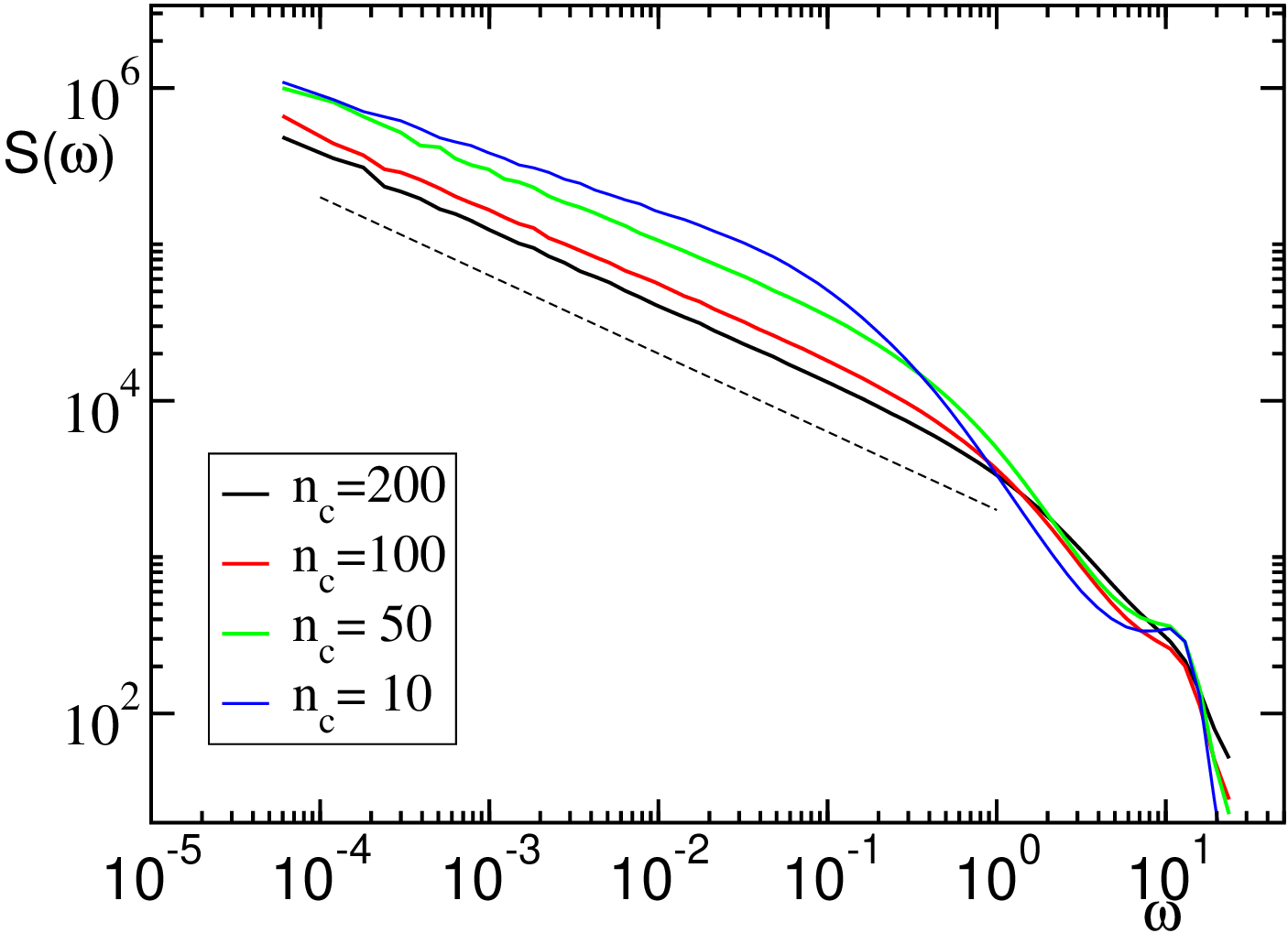}
\caption{Power spectrum of the heat flux for the stochastic
FPU model (cubic and quartic nonlinearities) with energy density 
$e=10$, $N=2048$ and different number
$n_c$ of colliding triplets. Units along the vertical axis are arbitrary.
Each curve is averaged over about 800 independent trajectories.
The thin dashed line corresponds to the law 
$\omega^{-1/2}$. } 
\label{fig2}
\end{center}
\end{figure}

\begin{figure}[ht]
\begin{center}
\includegraphics[clip,width=8cm]{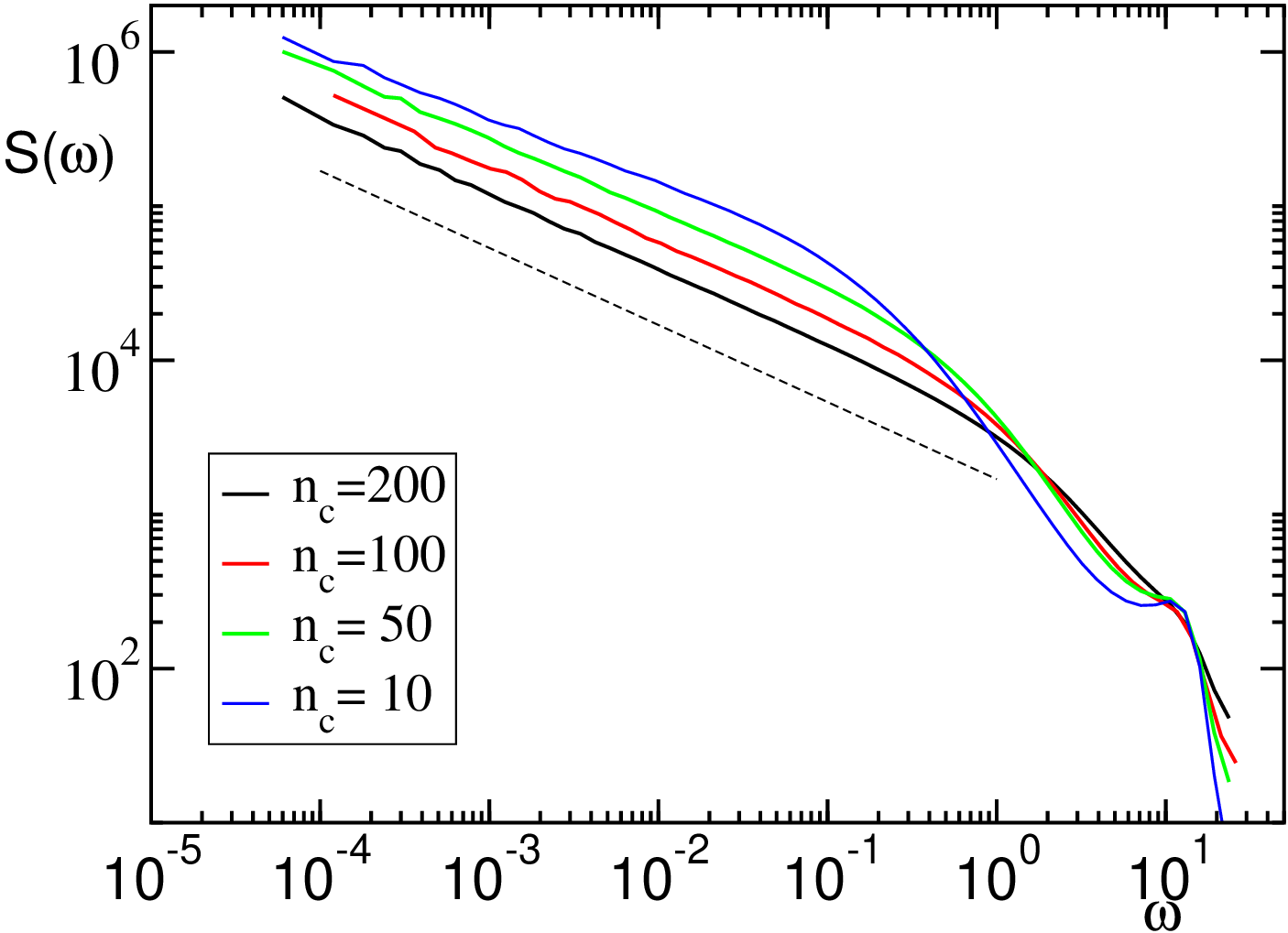}
\caption{Power spectrum of the heat flux for the stochastic 
$\beta$-FPU model (quartic nonlinearity) with energy density 
$e=10$, $N=2048$ and different number $n_c$ of colliding triplets.
Units along the vertical axis are arbitrary.
Each curve is averaged over about 800 independent trajectories.
The thin dashed line corresponds to the law 
$\omega^{-1/2}$. }
\label{fig3}
\end{center}
\end{figure}

Relying on these results, we have also studied the power spectra of the stochastic FPU
models (\ref{fpu}) with $g_2=g_3=g_4=1$ and 
$g_2=g_4=1$ and $g_3=0$ ($\beta$-FPU). For both models, we report the
power spectrum at fixed $N$ for different $n_c$ values (see Figs.~\ref{fig2},\ref{fig3}).
Analogously to what observed for the harmonic case, we see that upon increasing
$n_c$, the scaling region widens towards higher frequencies, thus allowing for
a more accurate analysis than in the purely deterministic case.
Moreover, in both cases the effective divergence exponent $\delta$ appears to
increase with $n_c$, approaching 1/2. More specifically, for the FPU with cubic and
quartic nonlinearities (Fig.~\ref{fig2}), $\delta$ increases from $\approx 0.35$ 
for $n_c=10$ to $\approx 0.48$ for $n_c=200$. Analogously, for the $\beta$-FPU case
(Fig.~\ref{fig3}), we find that $\delta$ increases from $\approx 0.41$, to
$\approx 0.47$.

Based on the existence of a few universality classes, a continuous dependence of
the  exponent $\delta$ on the number $n_c$ of collisions is quite unlikely.
However, one should recognize that the frequency scaling range is as wide as
three decades  and the dependence of $\delta$ on $\omega$ is pretty weak.
Altogether, the fact that $\delta$ is close to 1/2 for the larger $n_c$ values
suggests that the scaling behaviour predicted in \cite{olla} applies to a larger
class of systems and enforces the hypothesis of a second distinct universality
class. Moreover, it is interesting to notice that in both anharmonic models, the
divergence of thermal conductivity increases with the strength of the noise.
This means that, quite surprisingly, noise contributes to slowing down the decay
of energy current correlations.

\section*{Acknowledgments}
\medskip 

\small{
\noindent This work is supported by the PRIN2005 project {\it Transport
properties of classical and quantum systems} funded by MIUR-Italy, and
by ACI-NIM 168 {\it Transport Hors Equilibre} funded by Minist\'ere
Education Nationale, France. Part of the
numerical calculation were performed at CINECA supercomputing facility through
the project entitled {\it Transport and fluctuations in low-dimensional
systems}.
}


\begin{thebibliography}{00}

\bibitem{PR75} Y. Pomeau, R. R\'esibois, Phys. Rep. \textbf{19}  63 (1975).

\bibitem{kirk} T.R. Kirkpatrick, D. Belitz and J. V. Sengers,
J. Stat. Phys. {\bf 109}, 373 (2002).

\bibitem{DLQ89} D. D'Humieres, P. Lallemand, Y. Qian,
C. R. Acad. Sci. Paris {\bf 308}, serie II  585 (1989).

\bibitem{singlef} K. Hahn, J. K\"arger, and V. Kukla,
Phys. Rev. Lett. {\bf 76}, 2762 (1996).

\bibitem{MHSU86} D.T. Morelli, J. Heremans, M. Sakamoto, C. Uher,
Phys. Rev. Lett. \textbf{57}  869 (1986).

\bibitem{maruyama} S. Maruyama, Physica B \textbf{323}, 193 (2002).

\bibitem{LLP03} S. Lepri, R. Livi, A. Politi, Phys. Rep. {\bf 377}, 1 (2003).
[arXiv: cond-mat/0112193]


\bibitem{BLR00} F. Bonetto, J. Lebowitz, L. Rey-Bellet, in:
A. Fokas, A. Grigoryan, T. Kibble and B. Zegarlinski (Eds.), 
{\it  Mathematical Physics 2000}, Imperial College, London, 2000.

\bibitem{olla} G. Basile, C. Bernardin, and S. Olla
Phys. Rev. Lett. \textbf{96}, 204303 (2006).

\bibitem{PRV67} D.N. Payton, M. Rich, W.M. Visscher, Phys. Rev. \textbf{160}
706 (1967).

\bibitem{JPW68} E.A. Jackson, J.R. Pasta, J.F. Waters, J. Comput. Phys.
\textbf{2}  207 (1968).

\bibitem{N70} N. Nakazawa, Progr. Theor. Phys. Suppl. \textbf{45} 231 (1970).

\bibitem{LLP97} S. Lepri, R. Livi, A. Politi, Phys. Rev. Lett. {\bf 78}, 1896
(1997).

\bibitem{C86} G. Casati, Found. Phys. {\bf 16} 51 (1986).

\bibitem{H99} T. Hatano, Phys. Rev. E {\bf 59},  R1 (1999).

\bibitem{D01} A. Dhar, Phys. Rev. Lett. {\bf 86}  3554 (2001).

\bibitem{GNY02} P. Grassberger, W. Nadler, L. Yang, Phys. Rev. Lett. {\bf 89},
180601 (2002).

\bibitem{chaos} S. Lepri, R. Livi, A. Politi,
Chaos {\bf 15}, 015118 (2005)

\bibitem{LLP98} S. Lepri, R. Livi, A. Politi, 
Europhys. Lett. {\bf 43}, 271 (1998).

\bibitem{denis} P. Cipriani, S. Denisov and A. Politi,
Phys. Rev. Lett. {\bf 94}, 244301 (2005).

\bibitem{Bishop2} M. Bishop, J. Stat. Phys {\bf 29} (3), 623 (1982).

\bibitem{sandri} S. Lepri, P. Sandri, A. Politi
Eur. Phys. J. B {\bf 47}, 549 (2005).



\bibitem{CFVV84} G. Casati, J. Ford, F. Vivaldi, W.M. Visscher,
Phys. Rev. Lett. \textbf{52}  1861 (1984).

\bibitem{GH85} M.J. Gillan, R.W. Holloway, J. Phys. C \textbf{18} 5705 (1985).

\bibitem{HLZ98} B. Hu, B. Li, H. Zhao, Phys. Rev. E \textbf{57}  2992 (1998).

\bibitem{AK00} K. Aoki, D. Kusnezov, Phys. Lett. A \textbf{265}  250 (2000).

\bibitem{GLPV00} C. Giardin\'a, R. Livi, A. Politi, M. Vassalli, 
Phys. Rev. Lett.  2144 (2000) .

\bibitem{GS00} O. V. Gendelman, A. V. Savin, 
Phys. Rev. Lett. \textbf{84}  2381 (2000).

\bibitem{BO05} C. Bernardin, S. Olla, J. Stat. Phys. \textbf{121} 271,
(2005).

\bibitem{helfand} E. Helfand, Phys. Rev. {\bf 119}, 1 (1960).

\bibitem{klaft} S. Denisov, J. Klafter and M. Urbakh, Phys. Rev. Lett.
{\bf 91}, 194301 (2004).

\bibitem{blumen} A. Blumen, G. Zumofen, and J. Klafter, Phys. Rev. A {\bf 40},
3964 (1989).

\bibitem{forster} D. Forster, {\it Hydrodynamic Fluctuations, Broken symmetry
and Correlation Functions} (W. A. Benjamin, Reading, 1975).



\bibitem{KT} R. Kubo, M. Toda, and N. Hashitsume, {\it Statistical Physics II} 
Springer Series in Solid State Sciences, Vol. 31, Springer, Berlin, 1991.




\bibitem{EPJST}  L. Delfini, S. Denisov, S. Lepri, R. Livi, P. Mohanty, 
A. Politi, 
Eur. Phys. J. Special Topics \textbf{146}, 21 (2007)
[arXiv: cond-mat/0702212].

\bibitem{mct} W. G\"otze in 
\textit{Liquids, freezing and the glass transition},
edited by J. P. Hansen, D. Levesque and J. Zinn-Justin 
(North Holland, Amsterdam 1991); R. Schilling in 
\textit{Collective dynamics of nonlinear and 
disordered systems}, edited by G.Radons, W. Just and P. H\"aussler 
(Springer, Berlin, 2003).

\bibitem{SS97} J. Scheipers, W. Schirmacher, Z. Phys. B \textbf{103}, 547
(1997).

\bibitem{L98} S. Lepri, Phys. Rev. E {\bf 58} 7165 (1998).

\bibitem{DLLP06} L. Delfini, S. Lepri, R. Livi and A. Politi,
Phys. Rev. E \textbf{73}, 060201(R) (2006).

\bibitem{DLLP07} L. Delfini, S. Lepri, R.Livi, A. Politi, 
J. Stat. Mech. (2007) P02007. [arXiv: cond-mat/0611278]

\bibitem{B07} J.P. Bouchaud, arXiv:0705.0989v1 [cond-mat.dis-nn].

\bibitem{NR02}  O. Narayan, S. Ramaswamy, Phys. Rev. Lett.
{\bf 89},  200601 (2002).


\bibitem{P03} A. Pereverzev,
Phys. Rev. E \textbf{ 68}, 056124 (2003).

\bibitem{N07} B.G. Nickel, J. Phys. A: Math. Theor.  
\textbf{40} 1219 (2007). 

\bibitem{S07} J. Lukkarinen and H. Spohn, arXiv:0704.1607v1 [math-ph]

\bibitem{canadesi} G. R. Lee-Dadswell, B. G. Nickel, and C. G. Gray, 
Phys. Rev. E {\bf 72} 031202 (2005).

\bibitem{klages} S. Lepri, R. Livi, A. Politi in: R.Klages, G.Radons, 
I.M.Sokolov (Eds.), \textit{Anomalous Transport: Foundations and Applications}, 
Wiley-VCH, Weinheim (2007)

\bibitem{LLP03b} S. Lepri, R. Livi, A. Politi, 
Phys. Rev. E {\bf 68} 067102 (2003).

\bibitem{CP03} G. Casati and T. Prosen, Phys. Rev. E {\bf 67}, 015203(R)
(2003).

\bibitem{WL04} J. S. Wang and B. Li, Phys. Rev. Lett.
{\bf 92},  074302 (2004).





\end{thebibliography}
\end{document}